# Rapid Synthesis of Thermoelectric SnSe Thin Films by MPCVD

Yuyu Feng[a], Xi Zhang[a], Li Lei[b], Ya Nie[a], Gang Xiang[a] *

[a] College of Physics, Sichuan University, Chengdu, 610064, China

[b] Institute of Atomic and Molecular Physics, Sichuan University, Chengdu, 610064, China

**Abstract**

Microwave plasma chemical vapor deposition (MPCVD) has been traditionally used to synthesize carbon-based materials such as diamonds, carbon nanotubes and graphene. Here we report that a rapid and catalyst-free growth of SnSe thin films can be achieved by using single-mode MPCVD with appropriate source materials. The analysis combing microscope images, X-ray diffraction patterns and lattice vibration modes shows that the grown thin films were composed of orthorhombic structured SnSe polycrystals mainly along the (111) direction. Further thermoelectric (TE) characterizations reveal that the power factor of the SnSe films reached 3.98 $\mu W \cdot cm^{-1} K^{-2}$ at 600 K, comparable to the highest reported values of SnSe thin films. Our results may open an avenue for rapid synthesis of new types of materials such as IV-VI compounds and be useful for TE application of these materials.

*Corresponding author: gxiang@scu.edu

1. Introduction

Recently, IV-VI compound semiconductors have drawn a great deal of interest owing to their layered structures and outstanding physical properties. [1-7] Among them, SnSe compounds show good thermoelectric (TE) properties [2,8-10] and has been synthesized by various methods including thermal evaporation [11-13], electrodeposition [14], atomic layer deposition [15], solution process [16] and pulsed layer deposition [17]. Single crystalline SnSe is a layered orthorhombic structured compound semiconductor with the Pnma space group and usually experiences a phase transition from Pnma to a high symmetry Cmcm space group at around 800 K. Although the TE properties of SnSe single crystals are outstanding at high temperatures, they are usually poor below 800 K owing to the low carrier concentrations in SnSe single crystals with the Pnma space group.

To improve TE properties at temperatures below 800 K, thin film fabrication has been offered as an effective way within the context of utilizing textured microstructure and film continuity of SnSe polycrystals.[16] Furthermore, thin films are desirable for miniature devices such as micro TE generator and flexible TE sensors [3,10,13]. However, it is still a challenge to obtain high-quality SnSe thin films in a cheap and convenient way for further applications. On the other hand, microwave plasma chemical vapor deposition (MPCVD) has been employed as a fast and low-cost way to grow carbon-based materials, such as diamonds [19, 20], graphene [21, 22], carbon nanotubes [23, 24] and carbon nitrides [25, 26]. The biggest advantage of MPCVD is that the microwave plasma used in it can effectively heat up polar molecules to accelerate material synthesis. However, so far the study on the synthesis of new types of materials other than carbon-related or diamond-like ones by MPCVD is rare. Therefore, it will be desirable to employ MPCVD to establish a rapid, low-cost and easy way to synthesize SnSe thin films with good thermoelectric properties.

In this work, we employed MPCVD with appropriate source materials to grow high TE performance SnSe thin films. In order to easily control the microwave plasma, we used a single-mode MPCVD with a rectangular chamber to grow the samples. In this way, a rapid and catalyst-free synthesis of 15mm×15mm SnSe films were achieved. Further analysis reveals that the SnSe thin films were composed of orthorhombic structured SnSe polycrystals mainly along the (111) direction and exhibited good thermoelectric properties. The highest power factor of the SnSe thin films reached 3.98 $\mu W \cdot cm^{-1} K^{-2}$ at 600 K, comparable to the highest reported values of SnSe thin films.[16] The high TE performance of the MPCVD-synthesized SnSe thin films were caused by the highly-oriented texture, continuity and stoichiometric composition of the films[16], indicating a promising potential for applications of SnSe thin films.

## 2. Experimental methods

The $SeO_2$ (Sigma-Aldrich, 99.9%, 0.1mmol) powders and $SnCl_4 \cdot 5H_2O$ (Sigma-Aldrich, 98%, 0.1mmol) powders were chosen as the source materials. The critical temperatures for $SeO_2$ and $SnCl_4 \cdot 5H_2O$ to generate vapors are 315°C and 114.1°C, respectively, both of which are in the moderate range and can be conveniently reached by the MPCVD (Newman-Hueray, HMPS-2010SM).

Furthermore, SeO$_2$ has higher polarity than SnCl$_4$·5H$_2$O and therefore can absorb microwave energy more quickly and get heated more easily. As a result, SeO$_2$ and SnCl$_4$·5H$_2$O powders can be heated together to provide Se and Sn vapor sources simultaneously by the MPCVD. The microwave frequency of the MPCVD system is 2450 MHz, and it works in a single mode of TE$_{10}$. Two steps were involved in the growth. At the first step, the power of the microwave was kept at 500W to convert SnCl$_4$·5H$_2$O and SeO$_2$ powders into vapor rapidly. Then at the second step, the power was reduced to 120W to let SnSe deposit on the semi-insulating silicon (111) substrates. The pressure in the chamber was controlled at 700Pa with the forming gas (5% hydrogen and 95% argon) flow rate at 30sccm. The temperatures in the reaction chamber were about 380°C and 340°C for the two steps, respectively. In this way, the main product deposited on the substrates SnSe, and the other products (HCl and H$_2$O) were vaporized and removed from the chamber. It took only 18 minutes (3 minutes on the first stage and 15 minutes on the second stage) for producing 100nm thick SnSe films.

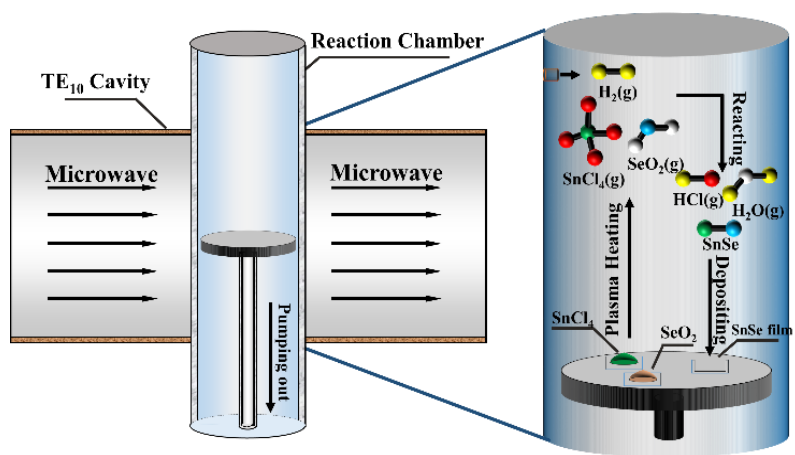

Figure 1. Synthesis process of the SnSe thin films

The schematic of the growth process in the chamber is shown in Fig. 1 and the reactions involved in the growth are listed as follows:

$$H_2 \rightarrow 2H^+ + 2e^- \qquad (1)$$

$$SnCl_4 \cdot 5H_2O \rightarrow Sn^{4+} + 4Cl^- + 5H_2O \qquad (2)$$

$$SeO_2 \rightarrow Se^{4+} + 2O^{2-} \qquad (3)$$

$$8H^+ + 4Cl^- + 2O^{2-} \rightarrow 4HCl + 2H_2O \qquad (4)$$

$$Sn^{4+} + Se^{4+} + 8e^- \rightarrow SnSe\downarrow \qquad (5).$$

At the beginning, hydrogen molecules were ionized to plasmas (Eq. 1), which had much higher chemical activity than normal hydrogen [27]. $SnCl_4 \cdot 5H_2O$ and $SeO_2$ were also ionized to plasmas at the same time (Eq. 2-3). Then H atoms in hydrogen plasmas were bonded with Cl atoms and O atoms, respectively (Eq. 4). Meanwhile, Sn (+4) atoms and Se (+4) atoms were reduced to Sn (+2) and Se (-2), respectively (Eq. 5) and then the product of SnSe was formed and deposited on the substrate. In this process, the appropriate amount of hydrogen was carefully used to guarantee that Sn atoms were reduced to +2 charge, so that our product was mainly SnSe but not $SnSe_2$ or Sn.

The morphologies of the SnSe thin films were characterized by atomic force microscope (AFM, Benyuan, CSPM-5500) and scanning electron microscopy (SEM, Thermo Scientific, Apreo-S). The elementary distribution of the SnSe film were investigated by SEM and energy dispersive X-ray spectrometry (EDS, Oxford, X-MaxN 80). The microstructures were characterized by high resolution transmission electron microscope (HRTEM, Talos, G2-F200S) and corresponding fast Fourier transformation (FFT). The crystal structures were analyzed by X-ray diffraction (XRD, Fangyuan, DX-2500) with a Cu-Kα (λ=0.15418nm) radiation source and the vibration modes of the SnSe lattice were characterized by Raman spectroscopy using custom-built confocal micro-Raman optical assembly with 532 nm excitation laser and an Andor EMCCD detector. Finally, the thermoelectric properties of the SnSe films were investigated.

3. Results and discussions

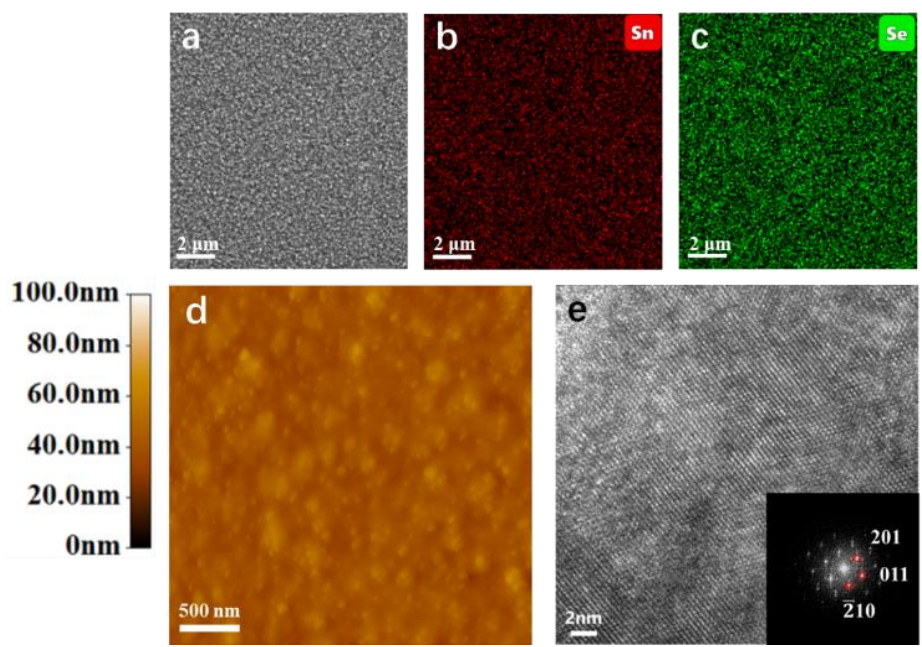

Figure 2. (a) SEM image of the SnSe film. (b) Sn and (c) Se elementary maps of the corresponding SnSe film (d) AFM image of the SnSe film with small circular protuberance textures. (e) HRTEM and FFT images of the SnSe film.

Fig. 2(a) shows the SEM image of a typical SnSe film and Fig. 2(b-c) shows the elementary maps of Sn and Se of the corresponding film. One can see that the elementary distribution of the SnSe film is uniform, and the stoichiometry of the Sn:Se films is about 1:1. Fig. 2(d) shows the AFM image of the SnSe film. Circular protuberance with diameters around 100 nm spread closely all over the sample surface, and the value of root mean square (RMS) of the surface is 4.78 nm. The HRTEM image of the SnSe film in Fig. 2(e) reveals a clear periodicity of the atomic arrangement in the SnSe crystal. The lower right part of Fig. 2(e) is the corresponding FFT pattern with the Miller indices (hkl) presented, which exhibits the clear crystallinity in the SnSe film.

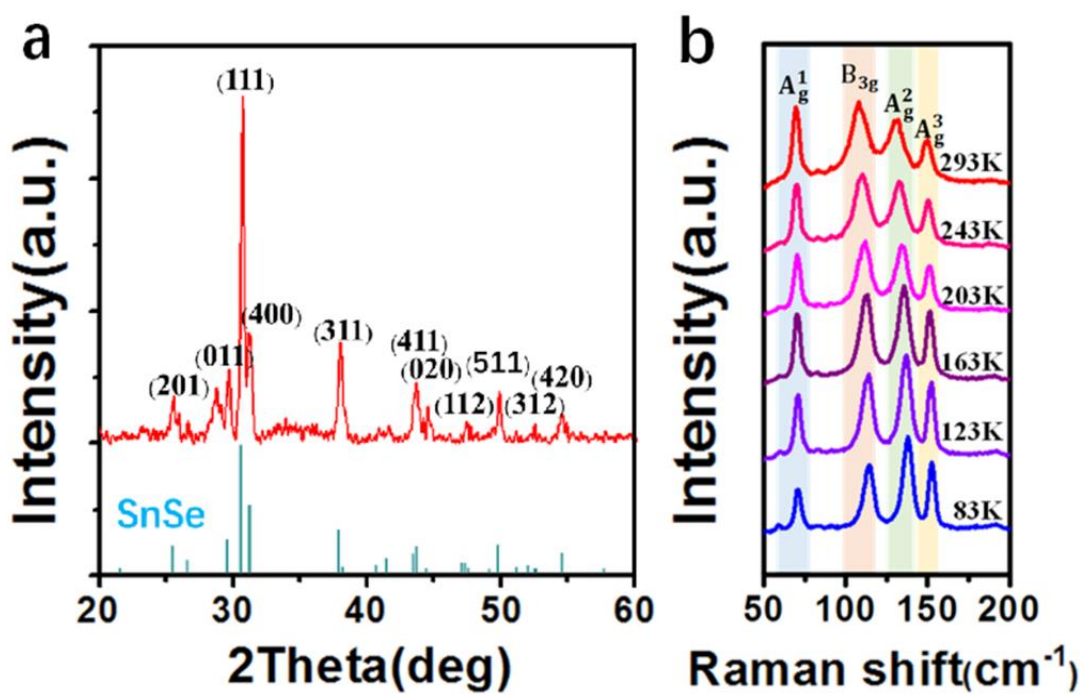

Figure 3. (a) XRD patterns of the SnSe film and the standard XRD patterns (ICDS: 89-0232) of SnSe. (b) Raman spectra of the SnSe film at different temperatures.

The XRD patterns of our SnSe films and the standard SnSe XRD patterns marked with corresponding Miller indices (hkl) are shown in Fig. 3(a). All the diffraction maxima in the spectrum can be indexed to the orthorhombic SnSe structure with space group Pnma (ICDS card No.89-0232), except the one at 2θ=28.65°, which belongs to the Si (111) substrate. The lattice constants were found to be $a$ = 11.502 Å, $b$ = 4.153 Å and $c$ = 4.450 Å, respectively. The patterns reflect that the (111) plane is the strongest orientation, indicating a highly oriented texture of the SnSe films. The XRD peaks of the SnSe films are sharp, among which the (111) peak exhibits a full width half maximum (FWHM) of only 0.206°, corresponding to a calculated crystalline dimension of about 46 nm.

In order to characterize the vibration modes of the lattice and the phase components in the samples, Raman spectroscopy of the SnSe films at different temperatures were investigated. As shown in Fig. 3(b), the spectra reveal four main vibrational modes at 70 cm$^{-1}$, 107 cm$^{-1}$, 130 cm$^{-1}$ and 150 cm$^{-1}$, in which the peak at 107 cm$^{-1}$ belongs to the B$_{3g}$ phonon mode and the other three belong to A$_g$ phono modes. A$_g$ and B$_{3g}$ are two rigid shear modes of a layer with respect to its neighbors in the $b$ and $c$ directions, respectively, which are the characteristic planar vibration modes of the lattice [28]. All the observed vibrational modes agree well with the characteristic modes of orthorhombic structured SnSe crystals, which is stable at ambient conditions [29]. As the temperature decreases, the Raman peaks become sharper and slightly blue-shifted, which are caused by lower thermal disturbance and reduction of stress at lower temperatures.

Then the TE properties of the polycrystalline SnSe films synthesized by single-mode MPCVD were investigated. As we know, the efficiency of TE materials is usually measured by TE power factor (S$^2\sigma$) and a dimensionless figure of merit

ZT=S²σT/($\kappa_L$+$\kappa_e$), where S is the Seebeck coefficient, σ the electrical conductivity, T the temperature, and $\kappa_L$ and $\kappa_e$ the photonic and electronic contributions to the thermal conductivity, respectively. As shown in Fig. 4(a), the maximal Seebeck coefficient of our SnSe thin films occurs at 600 K, which is 627.7 µV/K and comparable to those of high-quality SnSe crystalline bulks [8,9,30]. As the temperature increases, the electrical conductivity of the SnSe films first decreases then increases between 9.3 S/cm and 19.2 S/cm (Fig. 4(b)), while the thermal conductivity keeps decreasing from 0.92 to 0.58 W/(m·K), similar to the results observed in crystalline SnSe bulks [34]. As a result, the TE power factor of the SnSe films reached up to 3.98 µW·cm$^{-1}$K$^{-2}$ at 600 K, which is at least one order of magnitude higher than that of the SnSe films reported by Burton *et al.* very recently [13], higher than that of single crystalline SnSe bulks measured along the *b*-axis in the moderate temperature range (500 K – 650 K) [9], and comparable to the highest reported value of SnSe thin films at 550 K.[16] Furthermore, the ZT value of the MPCVD synthesized SnSe films reached maximal value of 0.335 at 600 K, which is higher than those of the SnSe films prepared by other sophisticated methods such as pulsed laser deposition [17], and comparable to those of the crystalline SnSe bulks [32-35].

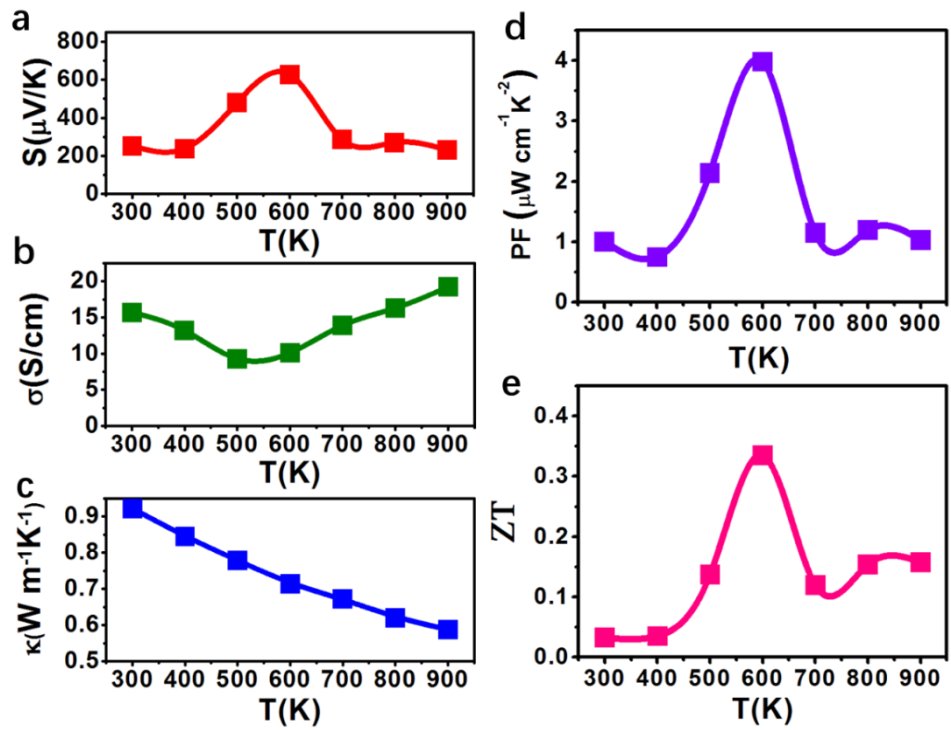

Figure 4. Thermoelectric properties of the SnSe films as a function of temperature. (a) Seebeck coefficient. (b) Electrical conductivity. (c) Total thermal conductivity. (d) Power factors. (e) ZT values.

## 4. Conclusions

In this work, we demonstrated a rapid and catalyst-free synthesis of SnSe thin films by single-mode MPCVD, extending the types of materials grown by MPCVD from traditional carbon-related materials to the IV-VI compounds. The grown films were composed of nanoscale SnSe polycrystals and exhibited high performance thermoelectric properties. The results may be useful for the investigation of MPCVD-grown new types of materials such as thermoelectric IV-VI compounds and the application of these material.

**Acknowledgement**

This work was supported by National Key R&D Program of China through Grant No. 2017YFB0405702 and by the Natural Science Foundation of China (NSFC) through Grant No. 51672179.